\documentclass[11pt]{article} 
\usepackage{hyperref} 
\pdfoutput=1 
\begin{document} 
\title{Holy balls!\\
	Entry\#: V040} 
\author{Michael Wright, Ken Langley, Jesse Belden$^\dagger$ \& Tadd Truscott \\
\\
Department of Mechanical Engineering \\ 
Brigham Young University, Provo, UT, 84602 USA \\
$^\dagger$Naval Undersea Warfare Center, Newport, RI, 02840 USA} 
\maketitle 
\begin{abstract} 
We demonstrate the behavior of three balls skipping off of the water surface: a Superball\textsuperscript{\textregistered}, a racquetball, and a water bouncing ball (Waboba\textsuperscript{\textsuperscript{\textregistered}}).  The three balls have rebound coefficients of 0.9, 0.8 and 0.2, respectively.  However, we notice that the Waboba\textsuperscript{\textregistered} bounces better than the others, but why?  The Superball\textsuperscript{\textregistered} has a high coefficient of restitution, creating large rebounds.  Here the impact is angled to the free surface, but the inelastic response and large mass ratio forces the ball underwater without skipping.  The racquetball has a lower mass ratio and a more elastic response to impacts.  Also thrown at a shallow angle, it bounces off of the surface of the water 1-3 times before coming to rest.  The Waboba\textsuperscript{\textregistered} flattens inside the cavity allowing it to skip off of the surface more easily.  The flattened ball looks more  like a skipping stone than a sphere due to its large elastic deformation at impact.  Examining the reaction of a skipping stone\cite{Clanet2004} we see that the stone creates a cavity in which it planes, slipping out of it with some upward velocity.  The Waboba\textsuperscript{\textregistered}  behaves like a skipping stone planing on the surface of the water, allowing it to bounce upwards of 20 times and traveling nearly 60 meters.   While some skill is needed to throw the Waboba\textsuperscript{\textregistered} across a pond, It adapts to each skip because of its elastic response, whereas the stone must be thrown perfectly in order to gain the best skipping advantage. 
\end{abstract} 
%
\bibliographystyle{abbrv}
\bibliography{bib}

\end{document}